\documentclass[12pt]{article}
\usepackage{amsmath,amssymb} 
\usepackage{subcaption}   

\usepackage{amsfonts}
\usepackage{enumerate}
\usepackage{ytableau}
\usepackage{hyperref}
\usepackage{cite}
\usepackage{bm}
\usepackage{CJK} 
\usepackage{tikz}
\usepackage{indentfirst}
\usepackage{blkarray}  
\usepackage{multirow}
\usepackage{booktabs}
\usepackage{amsmath}
\usepackage{array}
\usepackage{siunitx}
\usepackage{graphicx}
\usepackage{longtable}
\usepackage{xltabular}

\def\md{\mathrm{d}} 
\def\me{\mathrm{e}} 
\def\mi{\mathrm{i}} 
\def\nn{\nonumber} 
\def\({\left(} 
\def\){\right)} 
\def\[{\left[} 
\def\]{\right]}

\def\>{\rangle}
\def\<{\langle}

\begin{document}
	
	\title{\textbf{Half-wormholes in a complex SYK model}}
	\vspace{14mm}
	\author{Yingyu Yang\footnote{yangyingyu@pku.edu.cn}}
	\date{}
	\maketitle
	
	\begin{center}
		\it Center for High Energy Physics, Peking University, No.5 Yiheyuan Rd, Beijing 100871, P. R. China
		\vspace{10mm}
	\end{center}
	
	\begin{abstract}
		
	We  compute the half-wormhole contribution in a complex SYK model with one time point by exact computations and saddle point analysis. The chemical potential significantly influences the computation and the relative magnitude of the half-wormhole. When the chemical potential is large compared to the random coupling, the disk dominates  so that there is no wormhole and half-wormhole. When  the chemical potential is relatively small we find out the wormhole and the half-wormhole contributions.
	\end{abstract}
	
	\newpage
	\tableofcontents

\section{Introduction}

Recently the spacetime wormhole in the semi-classical computation has important applications in explaining many phenomena, such as the transition of the  page curve \cite{Almheiri:2019qdq,Penington:2019kki}, the late time behavior of the spectral form factor \cite{Saad:2019lba,Saad:2018bqo} and the  correlation function \cite{Saad:2019pqd}. However, including spacetime wormholes in the AdS/CFT correspondence \cite{Maldacena:1997re,Witten:1998qj,Gubser:1998bc}, which is believed to provide a non-perturbative definition of quantum gravity, leads to a contradiction called the factorization problem \cite{Maldacena:2004rf}. On the field theory side the partition function $Z_{LR}$ of two decoupled field theories can be factorized into the product of two individual partition functions of the subsystems $Z_{L}Z_{R}$. However on the gravity theory side the  wormhole connecting the two boundaries provides an additional contribution which leads to  $Z_{LR}\neq Z_{L}Z_{R}$, which is contradictory to the field theory side. This factorization problem can be avoided by introducing ensemble averages into the systems, it is natural that the averaged partition function does not factorize $\langle Z_{LR}\rangle\neq\langle Z_{L}\rangle \langle Z_{R}\rangle$. The thought that  wormholes relate to ensemble average dates back to the 1980s \cite{Coleman:1988cy,Giddings:1988cx,Giddings:1988wv}, which implies a possible conjecture between a bulk gravity theory and an ensemble of boundary field theories.  One famous such duality  is between the two-dimensional Jackiw-Teitelboim (JT) gravity \cite{Jackiw:1984je,Teitelboim:1983ux}  and the Sachdev-Ye-Kitaev (SYK) model \cite{Sachdev:1992fk,KitaevTalk2,Maldacena:2016hyu}. However in standard AdS/CFT correspondence the boundary theory is not in ensemble average, there is still a demand to consider the factorization problem in the existence of wormholes.  In \cite{Saad:2021rcu} the authors study this problem carefully.  They introduce a new kind of saddle called half-wormhole and propose that the wormhole plus the half-wormhole can restore the factorization. The analysis is explicitly done in a 0-dimensional SYK model or SYK model with one time point. Following this development there is research on half-wormholes in other models, such as  0-dimensional SYK model in different ensembles \cite{Peng:2022pfa} or Brownian SYK model \cite{Peng:2022pfa}, supersymmetric SYK model \cite{Forste:2024nsw} and some discussions about 1-dimensional SYK model and random matrix theory \cite{Tian:2022zpc}. Further work about half-wormholes can be found in \cite{Saad:2021uzi,Mukhametzhanov:2021nea,Garcia-Garcia:2021squ,Choudhury:2021nal,Mukhametzhanov:2021hdi,Goto:2021mbt,Blommaert:2021fob,Goto:2021wfs,Cheng:2022nra,Peng:2021vhs}.

In this paper we  compute the half-wormhole contribution in a complex SYK model \cite{Gu:2019jub,Davison:2016ngz} with one time point or 0-dimensional complex SYK model. The complex SYK model has more degrees of freedom   and richer dynamics to the real version, it is proposed in \cite{Gaikwad:2018dfc}  the gravity dual of a complex SYK model is JT gravity coupled to a Maxwell field.  Other related work can be found in \cite{Chaturvedi:2020jyy,Afshar:2019axx,Godet:2020xpk,Zhang:2025kty}. The chemical potential $\mu$ in  the complex SYK model is related to the stength of the bulk gauge field \cite{Gaikwad:2018dfc}, our study may suggest that the stength of the gauge field can affect the bulk configurations.     We first evaluate the averaged partition functions by exact computations of the Grassmann integrals, then perform the saddle point analysis with small $\mu$ and observe consistency between the two approaches. When  $\mu$ is large compared to the random coupling, the disk dominates so that there is no wormhole and half-wormhole. When  $\mu$ is very small, we can construct the half-wormhole following the same procedure in \cite{Saad:2021rcu,Peng:2022pfa}.

\section{0-dimensional complex SYK model}

The Hamiltonian of the 0-dimensional complex SYK model is 
\begin{align}
	H=\sum_{\substack{j_{1}<\dots<j_{q/2}\\k_{1}<\dots<k_{q/2}}}J_{j_{1}\dots j_{q/2},k_{1}\dots k_{q/2}}\psi^{\dagger}_{j_{1}}\dots \psi^{\dagger}_{j_{q/2}} \psi_{k_{1}}\dots \psi_{k_{q/2}}, \label{hamiltonian}
\end{align}
where  $\psi^{\dagger}_{i},\psi_{i}$ are complex Grassmann numbers and  the  couplings $J_{j_{1}\dots j_{q/2},k_{1}\dots k_{q/2}}$ are drawn from a complex Gaussian distribution with zero mean and the variance 
\begin{align}
	\langle |J_{j_{1}\dots j_{q/2},k_{1}\dots k_{q/2}}|^{2}\rangle=J^{2}\frac{(q/2)!(q/2-1)!}{N^{q-1}}.
\end{align}
Furthermore  $J_{j_{1}\dots j_{q/2},k_{1}\dots k_{q/2}}$ is anti-symmetric within both the $j$ and $k$ index sets seperately and  $(j_{1},\dots,j_{q/2})<(k_{1},\dots,k_{q/2})$.
The Hermiticity requirement of the Hamiltonian imposes an additional constraint on the couplings
\begin{align}
	J_{j_{1}\dots j_{q/2},k_{1}\dots k_{q/2}}=J^{\ast}_{k_{1}\dots k_{q/2},j_{1}\dots j_{q/2}}.
\end{align}
The path integral of the 0-dimensional complex SYK model is actually a complex Grassmann integral
\begin{align}
	z&=\int \md\psi \md\psi^{\dagger} \exp\left( \mu\psi^{\dagger}_{i}\psi_{i} -J_{j_{1}\dots j_{q/2},k_{1}\dots k_{q/2}}\psi^{\dagger}_{j_{1}}\dots \psi^{\dagger}_{j_{q/2}} \psi_{k_{1}}\dots \psi_{k_{q/2}}\right).
\end{align}
After ensemble average over the random couplings and introducing the $G,\Sigma$ fields by inserting the identity 
\begin{align}
	1=\int_{\mathbb{R}\times \mi\mathbb{R} } \frac{\md G\md \Sigma}{2\pi\mi/N} \exp\left[ -N\Sigma\left(G-\frac{1}{N}\sum_{i=1}^{N}\psi_{i}^{\dagger}\psi_{i}\right)\right],
\end{align}
we integrate out $\psi^{\dagger},\psi$ fields to obtain
\begin{align}
	\langle z\rangle&=\int_{\mathbb{R}\times \mi\mathbb{R} }\frac{\md G\md \Sigma}{2\pi\mi/N} \exp\left[ N\log \left(\mu+\Sigma \right) -N  \Sigma G+ \frac{N\bar{J}}{q}  G^{q}  \right], \label{muzaverageaction}
\end{align}
where we define $\bar{J}\equiv J^{2}(-1)^{q/2}$ for simplicity.

\section{Averaged theories}
Under ensemble average the factorization fails and  the remaining  contributions are disks or wormholes \cite{Saad:2021rcu,Peng:2022pfa}. We will systematically identify these contributions within this simple model.

\subsection{$\langle z\rangle$}

\begin{figure}[htbp]
	\centering
	\subcaptionbox{}{\includegraphics[scale=0.5]{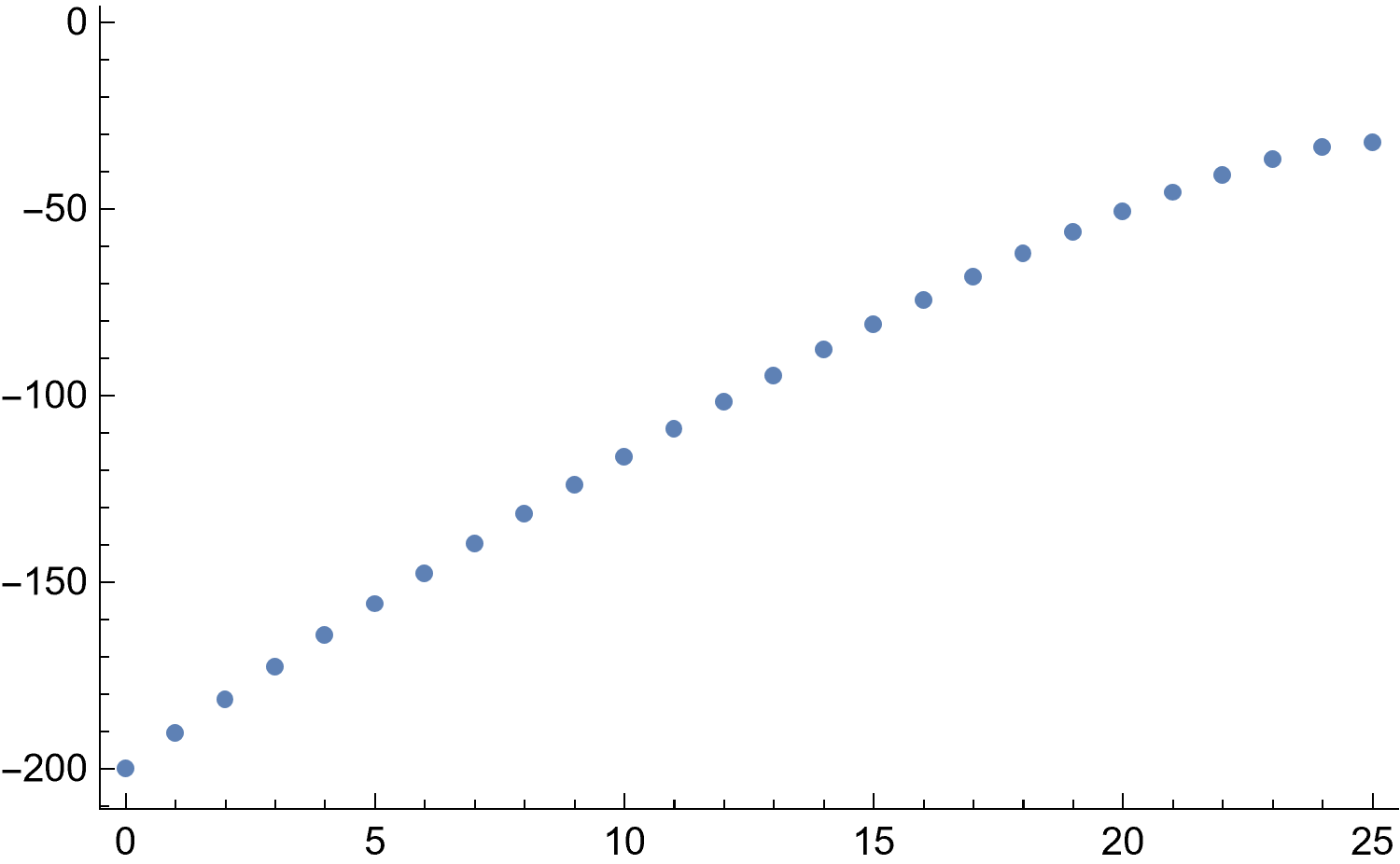}}
	\hfill
	\subcaptionbox{}{\includegraphics[scale=0.5]{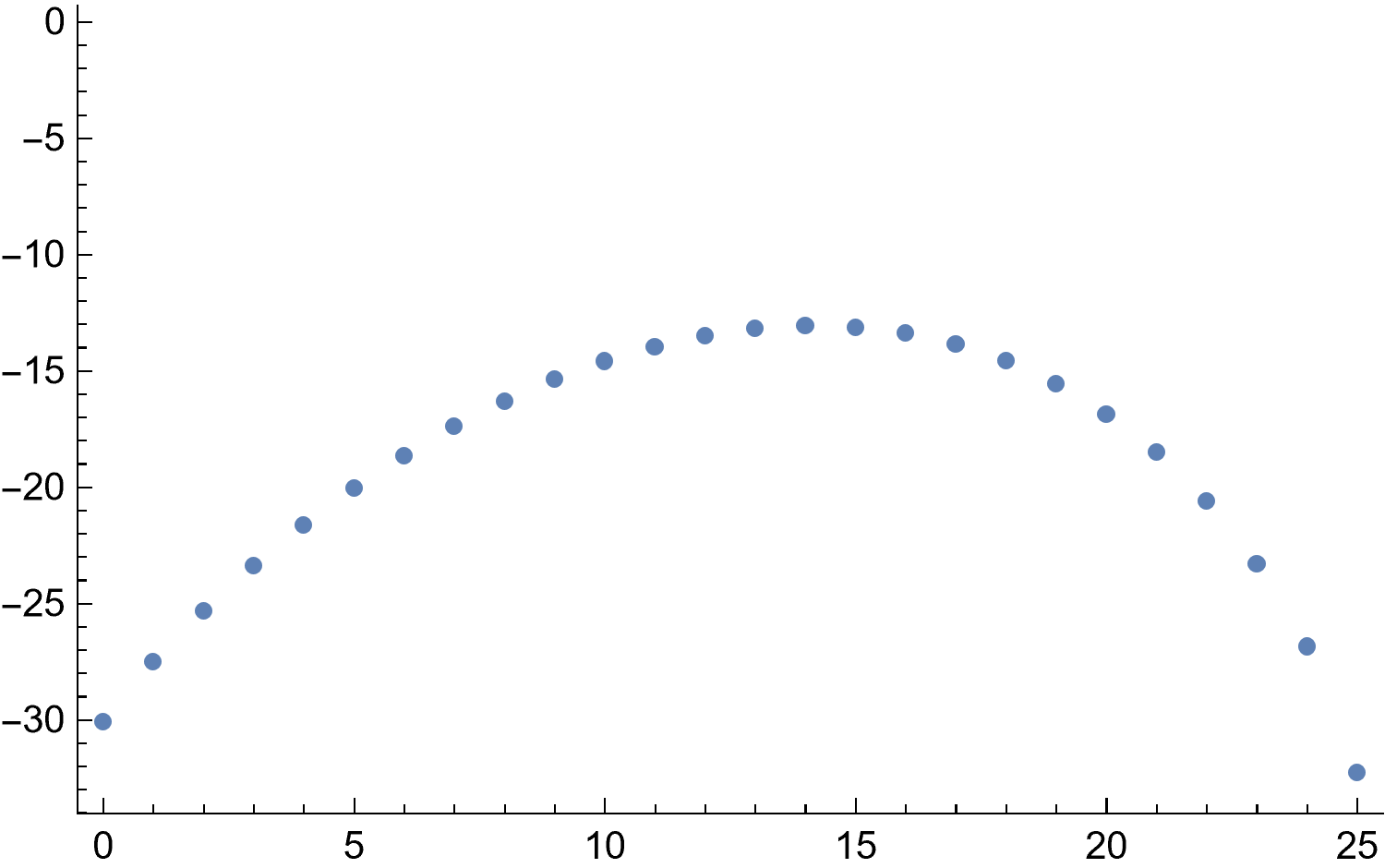}}
	\hfill
	\subcaptionbox{}{\includegraphics[scale=0.5]{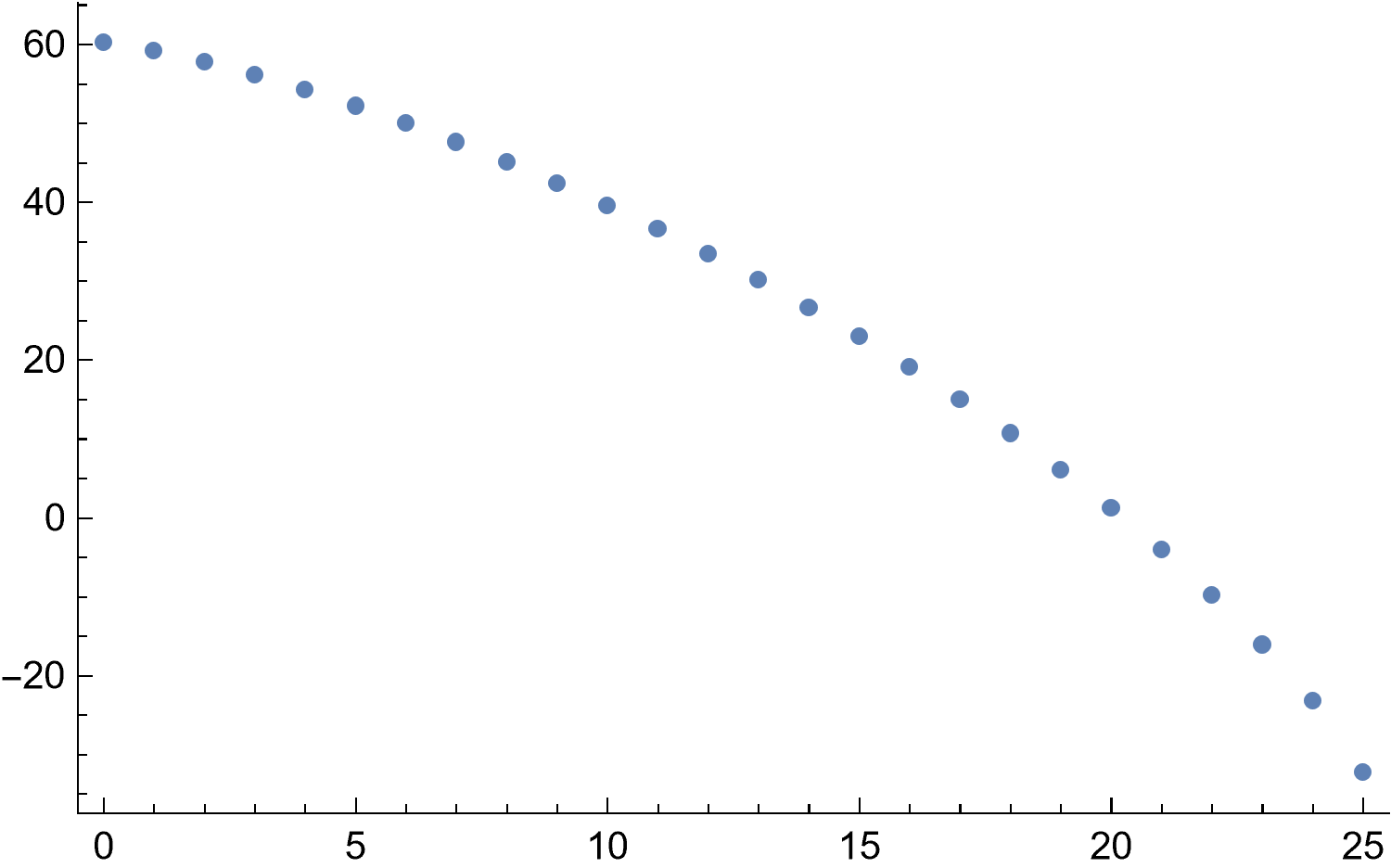}}
	\caption{The terms in the common logarithm $\log_{10}$ in  \eqref{muzaverage}  with  $n$ as the horizontal axis and   $N=100,q=4,\bar{J}=1$. (a) $\mu=0.01$. (b) $\mu=0.5$. (c) $\mu=4$.}
	\label{zdistribut}
\end{figure}

Following the procedure in \cite{Saad:2021rcu}  the averaged partition function can be expressed as
\begin{align}
	\langle z\rangle&=(\mu+N^{-1}\partial_{G})^{N}\me^{\frac{N\bar{J}}{q}G^{q}}|_{G=0},\\
	&=\sum_{m+nq=N,m,n\geq 0}\binom{N}{m}\mu^{m}N^{-nq}\left(\frac{N\bar{J}}{q}\right)^{n}\frac{(nq)!}{n!}, \label{muzaverage}
\end{align}
where $m,n$ are integers.
As shown in Fig. \ref{zdistribut} when $\mu$ is very small compared to $\bar{J}$ the dominant part will be $n=N/q$, while when $\mu$ is large compared to $\bar{J}$ the dominant part will be $n=0$ or  $\mu^{N}$. For intermediate $\mu$  different $n$'s  become dominant, therefore  only some of the terms in the summation dominate for specific $\bar{J},\mu$.

\subsection{$\langle z^{2}\rangle$}

\begin{figure}[htbp]
	\centering
	\subcaptionbox{}{\includegraphics[scale=0.5]{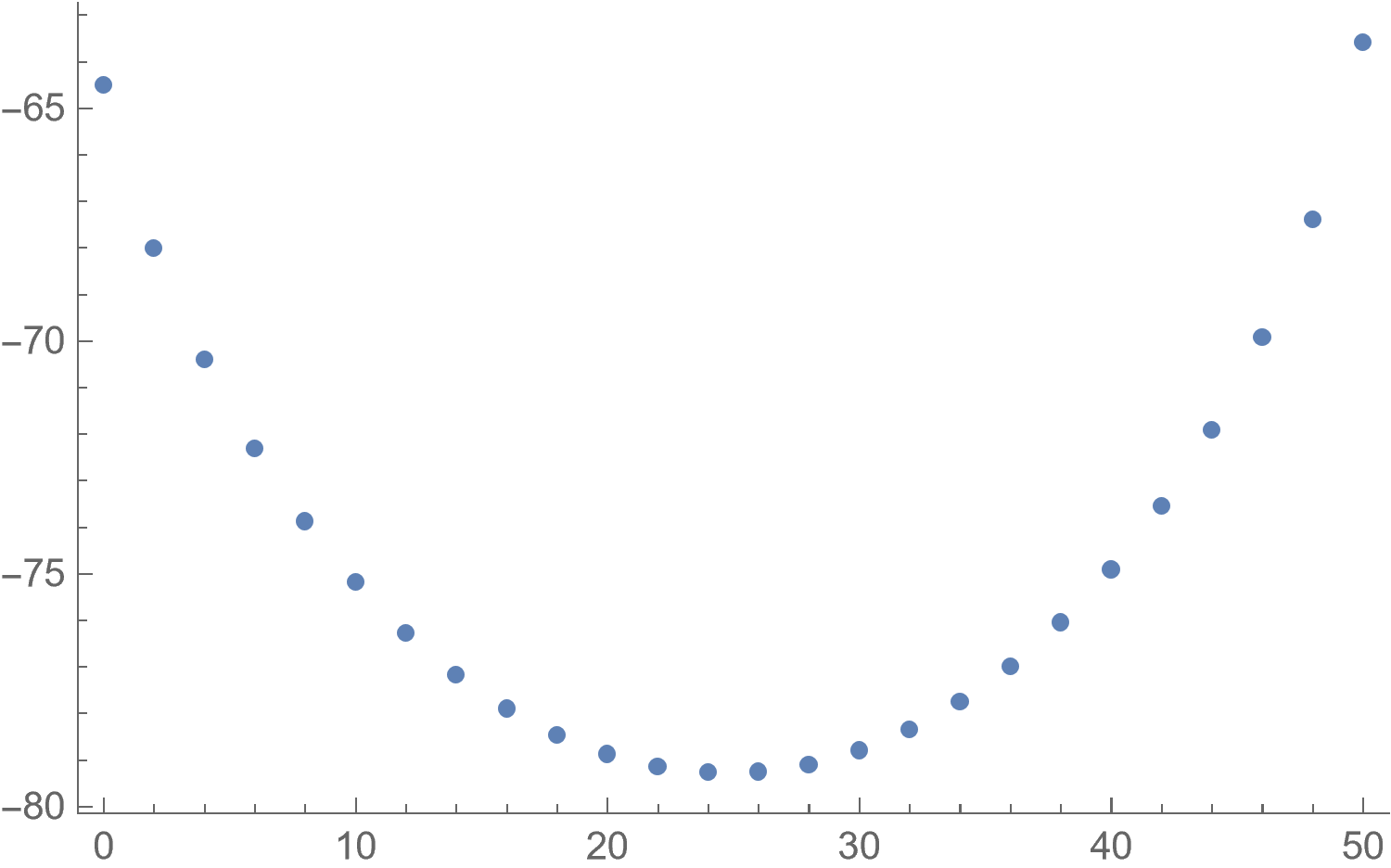}}
	\hfill
	\subcaptionbox{}{\includegraphics[scale=0.5]{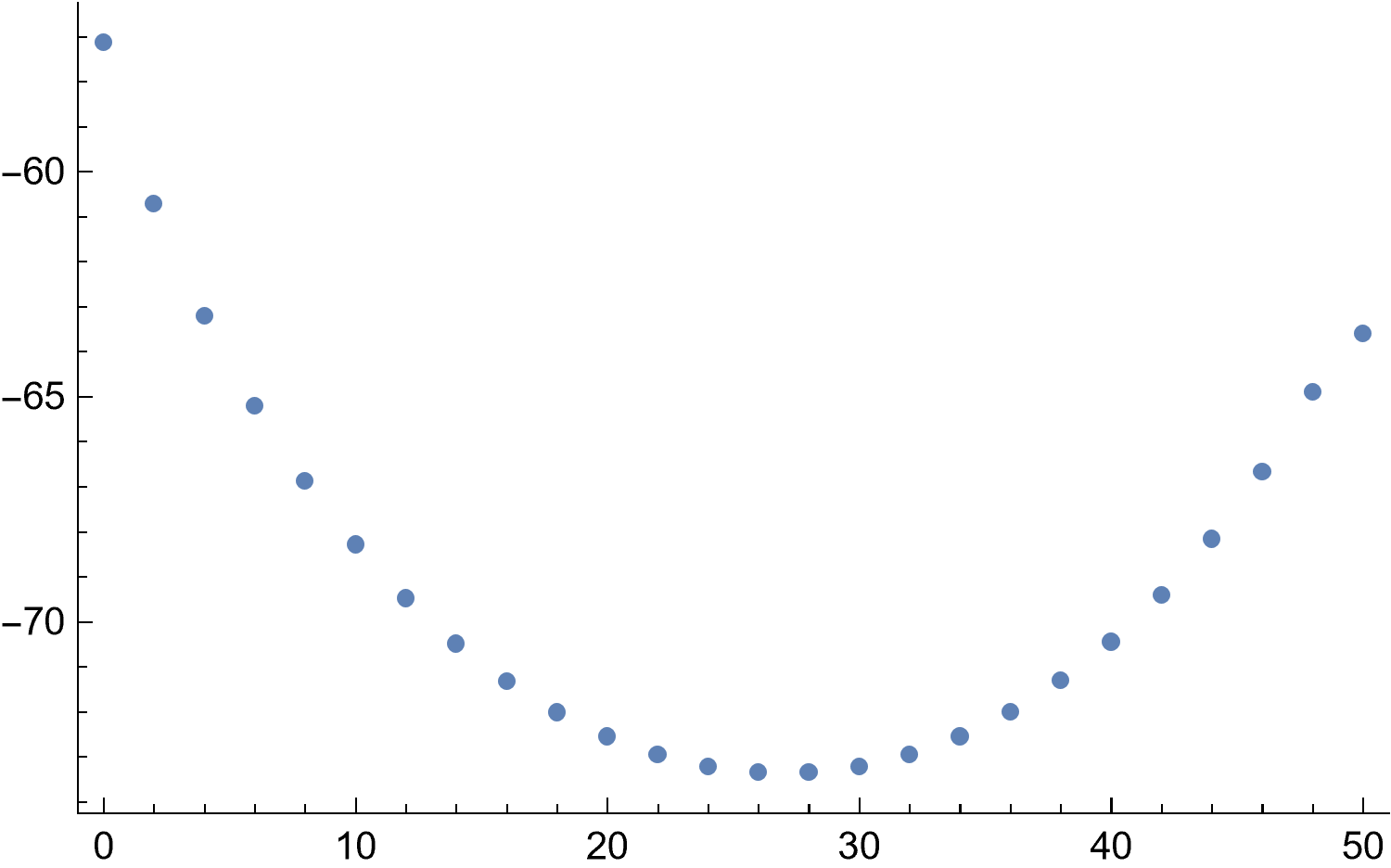}}
	\hfill
	\subcaptionbox{}{\includegraphics[scale=0.5]{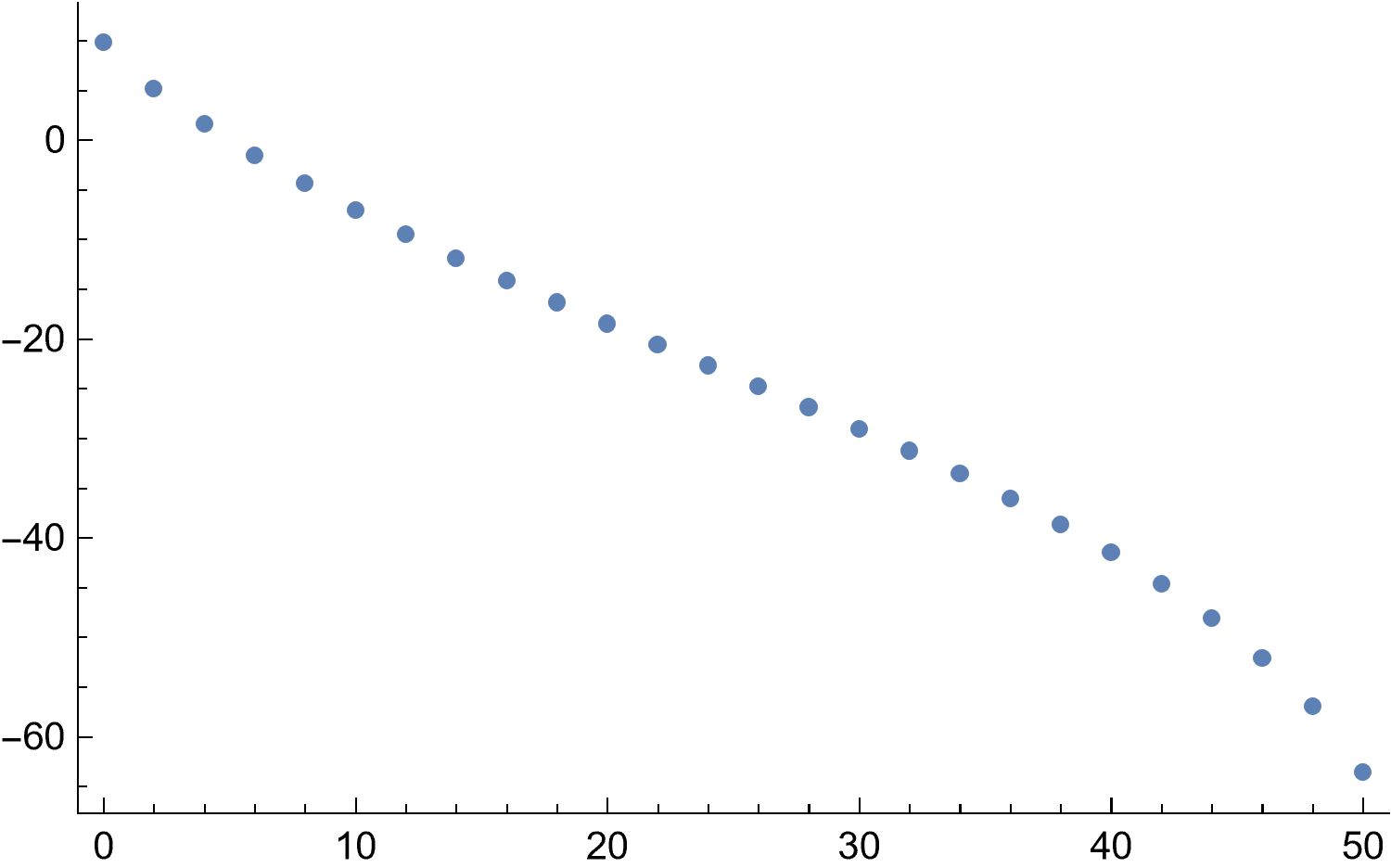}}
	\hfill
	\subcaptionbox{}{\includegraphics[scale=0.5]{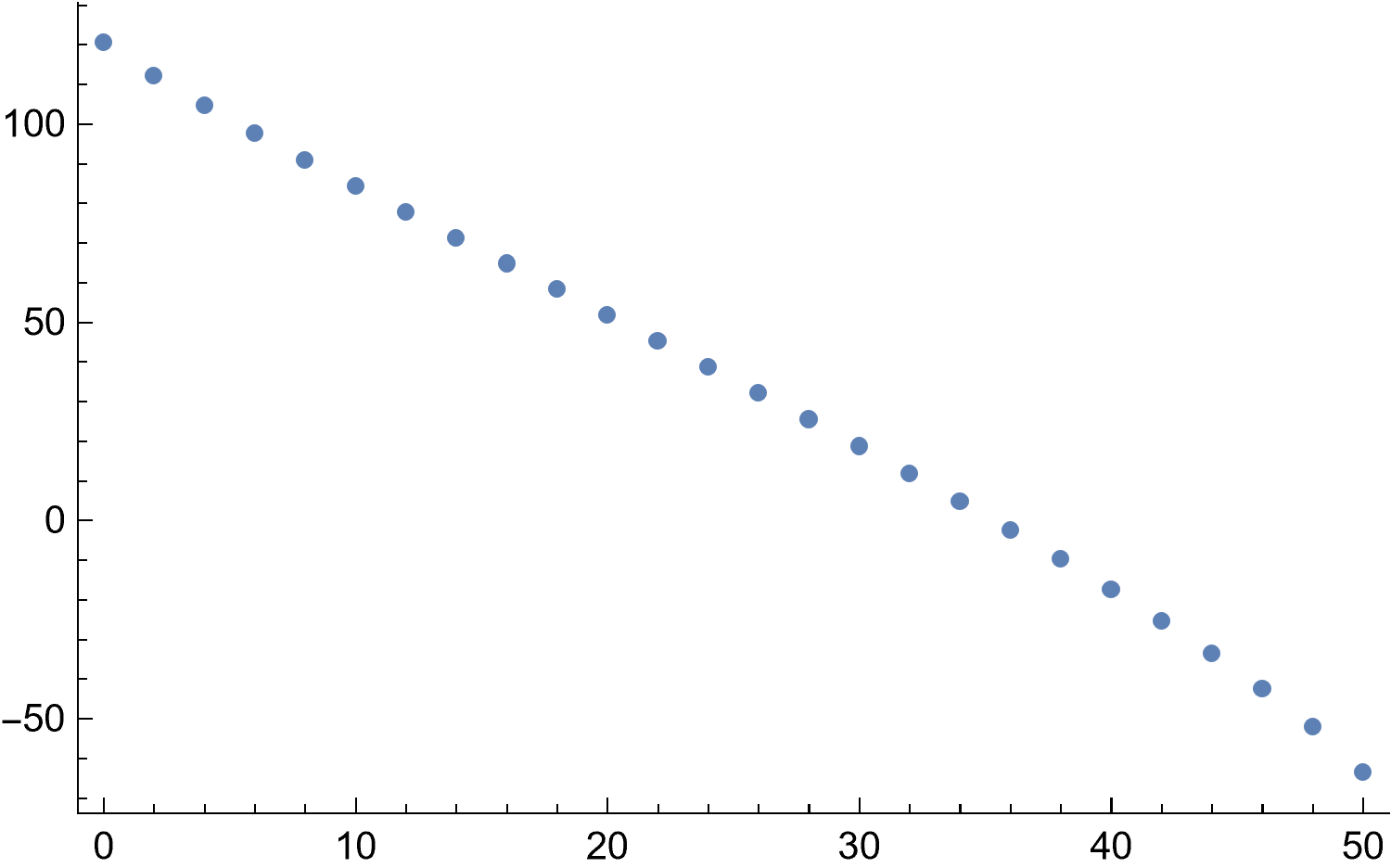}}
	\caption{The terms in the common logarithm $\log_{10}$ in  \eqref{muz2average} with $n_{3}$ as  the horizontal axis   and  $N=100,q=4,\bar{J}=1$. (a) $\mu=0.01$. (b) $\mu=0.1$. (c) $\mu=1$. (d) $\mu=4$.}
	\label{z2distribut}
\end{figure}

The averaged partition function of two copies of the 0-dimensional complex SYK model can be written as 
\begin{align}
	\langle z^{2}\rangle	&=\int_{\mathbb{R}\times \mi\mathbb{R} }  \frac{\md G_{ab}\md \Sigma_{ab}}{(2\pi\mi/N)^{4}} \left(\det \left(\mu\delta_{ab}+\Sigma_{ab} \right) \right)^{N} \exp\left[ -N  \Sigma_{ab} G_{ab}+ \frac{N\bar{J}}{q}  G_{ab}^{q/2}G_{ba}^{q/2}  \right], \label{z2averagemu}
\end{align}
where $a,b=L,R$.
Applying the binomial expansion twice to the determinant  yields
\begin{align}
	\langle z^{2}\rangle &=\sum_{\substack{n_{1}+n_{3}/2\leq \frac{N}{q},n_{i}\geq 0\\n_{2}+n_{3}/2\leq \frac{N}{q}}} \binom{N}{N- \frac{n_{3}q}{2}} \binom{N-\frac{n_{3}q}{2}}{N-\(n_{1}+\frac{n_{3}}{2}\)q}   \binom{N-\frac{n_{3}q}{2}}{N-\(n_{2}+\frac{n_{3}}{2}\)q} \mu^{2N-(n_{1}+n_{2}+n_{3})q} (-1)^{\frac{n_{3}q}{2}} \nn\\
	&\quad \times  N^{-(n_{1}+n_{2}+n_{3})q}  \left( \frac{N\bar{J}}{q} \right)^{n_{1}+n_{2}+n_{3}} \frac{2^{n_{3}}(n_{1}q)!(n_{2}q)!\left(\(\frac{n_{3}q}{2}\) !\right)^{2}}{n_{1}!n_{2}!n_{3}!}, \label{muz2average}
\end{align}
where $n_{1},n_{2},n_{3}$ are integers.
As shown in Fig. \ref{z2distribut} it seems that there are only two cases that either $n_{3}=2N/q$ is comparable to $n_{3}=0$ or not. When $\mu$ is large compared to $\bar{J}$ the dominant term will be $n_{3}=0$ which yields  $\langle z^{2}\rangle \approx \langle z\rangle^{2}\approx \mu^{2N}$.  When $\mu$ is very small compared to $\bar{J}$, there are two dominant terms $n_{3}=0,2N/q$ 
\begin{align}
	\langle z\rangle^{2},\quad  \frac{(N!)^{2}}{N^{2N}}\left(\frac{N\bar{J}}{q} \right)^{\frac{2N}{q}}  \frac{2^{\frac{2N}{q}}}{\frac{2N}{q}!}.  \label{z2dominate}
\end{align}
We will denote the two contributions as 
$\langle z\rangle^{2}$  and $ \langle \Phi^{2}(\mi\me^{\mi\pi/q}\mu) \rangle$, where $\langle z\rangle^{2}$  is from the direct computation \eqref{muzaverage}.  While   $ \langle \Phi^{2}(\mi\me^{\mi\pi/q}\mu) \rangle $  comes from  $ \langle \Phi^{2}(\sigma) \rangle $ defined in \eqref{muphi2average} by letting $\sigma=\mi\me^{\mi\pi/q}\mu$ and we will see it soon. 
Then we can have the approximation in  small $\mu$ 
\begin{align}
	\langle z^{2}\rangle \approx \langle z\rangle^{2}+\langle \Phi^{2}(\mi\me^{\mi\pi/q}\mu) \rangle. \label{muz2approxzp}
\end{align}

\subsection{$\langle z^{4}\rangle$}

Following the same procedure we derive the partition function for the averaged four copies of the theory
\begin{align}
	\langle z^{4}\rangle&=\int_{\mathbb{R}\times \mi\mathbb{R} }\frac{\md G_{ab}\md \Sigma_{ab}}{(2\pi\mi/N)^{16}} \exp\left[ N\log\det \left(\mu\delta_{ab}+\Sigma_{ab} \right) -N  \Sigma_{ab} G_{ab}+ \frac{N\bar{J}}{q}  G_{ab}^{q/2}G_{ba}^{q/2}  \right], \label{muz4average}
\end{align}
where $a,b=L,R,\bar{L},\bar{R}$.
The computation is very cumbersome, but remains tractable if we only care about the dominant terms. Explicitly we have 
\begin{align}
	\langle z^{4}\rangle&=\left(\det\left( \mu\delta_{ab} +N^{-1}\partial_{G_{ab}} \right)\right)^{N} \exp\left[ \frac{N\bar{J}}{q}  G_{ab}^{q/2}G_{ba}^{q/2} \right] \vert_{G_{ab}=0}, \label{z4aveaction}
\end{align}
 the matrix of the derivative can be written as  
\begin{align}
	\left(
	\begin{array}{cccc}
	\mu+\partial_{G_{LL}} & \partial_{G_{LR}} & \partial_{G_{L\bar{L}}} & \partial_{G_{L\bar{R}}}\\
		\partial_{G_{RL}} &\mu+\partial_{G_{RR}} & \partial_{G_{R\bar{L}}}&\partial_{G_{R\bar{R}}}\\
		\partial_{G_{\bar{L}L}} &\partial_{G_{\bar{L}R}} & \mu+\partial_{G_{\bar{L}\bar{L}}}& \partial_{G_{\bar{L}\bar{R}}}\\
		\partial_{G_{\bar{R}L}} &\partial_{G_{\bar{R}R}} & \partial_{G_{\bar{R}\bar{L}}}& \mu+\partial_{G_{\bar{R}\bar{R}}}\\
	\end{array}
	\right) , \label{muz4matrix}
\end{align}
where we omit the factor $N$ for simplicity.
Whether a term is dominant or not depends on the values of $\bar{J},\mu$. When  $\mu$ is large compared to $\bar{J}$  we obtain  $	\langle z^{4}\rangle \approx   \langle z\rangle^{4}\approx \mu^{4N}$. While when $\mu$ is very small compared to $\bar{J}$ we can  find the dominant terms from the matrix of the derivative  \eqref{muz4matrix}.  We can see it from the computation of $\< z^{2}\>$, whose dominant terms in small $\mu$ come from the below derivatives
\begin{align}
	\(\mu+\partial_{G_{LL}}\)^{N}\(\mu+ \partial_{G_{RR}}\)^{N},\quad \partial_{G_{LR}}^{N}\partial_{G_{RL}}^{N}. \label{z2avedominant}
\end{align}
Therefore we can get the dominant terms of $\<z^{4}\>$ from the matrix  \eqref{muz4matrix} by finding the products of \eqref{z2avedominant}, then we have
\begin{align}
	\langle z^{4}\rangle \approx   \langle z\rangle^{4}+6 \langle z\rangle^{2} \langle \Phi^{2}(\mi\me^{\mi\pi/q}\mu) \rangle + 3\langle \Phi^{2}(\mi\me^{\mi\pi/q}\mu) \rangle^{2}, \label{muz4approxzp}
\end{align}
where $ \langle \Phi^{2}(\mi\me^{\mi\pi/q}\mu) \rangle $ denotes the second term in \eqref{z2dominate}.

\section{Saddle point analysis}

For the averaged theories  we can also take the saddle point analysis.  For general $\mu,q$ the saddle point equations do not have analytic solutions, therefore our computation is the small $\mu$ perturbation to the saddles in the case of $\mu=0$ like in \cite{Saad:2021rcu}. 

\subsection{$\langle z\rangle$} 

We start with the partition function \eqref{muzaverageaction} by the redefinition to make the integral more convergent
\begin{align}
	\Sigma=\mi \me^{-\frac{\mi\pi}{q}}\sigma,\quad G=\me^{\frac{\mi\pi}{q}}g, \label{gsredefinition}
\end{align}
after varying over $g,\sigma$ we get the saddle point equations 
\begin{align}
	\(\mi\sigma+\mu\me^{\frac{\mi\pi}{q}}\) g=1,\quad  \mi \sigma +\bar{J}g^{q-1}=0. \label{zaverageeq}
\end{align}
The two equations lead to 
\begin{align}
	\bar{J} g^{q}-\mu g \me^{\frac{\mi\pi}{q}}+1=0,
\end{align}
which lacks a general analytic solution.  We can perform the small $\mu$ perturbation analysis  with the following ansatz 
\begin{align}
	g=g_{0}+g_{1},
\end{align}
where $g_{0}\sim O(1)$ and $g_{1}\sim O(\mu)$.  Substituting them into the equation we obtain the solution 
\begin{align}
&	g_{0}=\bar{J}^{-\frac{1}{q}}\me^{\frac{\mi\pi (2m+1)}{q}},\quad m=0,1,\dots q-1,\\
&	 g_{1}=-\frac{1}{q}\mu \me^{\frac{\mi\pi}{q}}g_{0}^{2},
\end{align}
which are identified as disks.
Inserting the solution into the path integral we have 
\begin{align}
	&\frac{1}{\sqrt{q}}\bar{J}^{\frac{N}{q}}\me^{-\(1-\frac{1}{q}\)N+N\mu\bar{J}^{-\frac{1}{q}}\me^{\frac{\mi\pi (2m+2)}{q}}},  \label{zaveragesaddle}
\end{align}
the contributions of the saddles are different. We can also implement the small $\mu$ perturbation in the exact computation \eqref{muzaverage}. Explicitly we only keep $n=\frac{N}{q},\frac{N}{q}-1$ in the summation and  apply Stirling's formula, then we have
\begin{align}
	\langle z\rangle &\approx \sqrt{q} \bar{J}^{\frac{N}{q}} \me^{-\left( 1-\frac{1}{q} \right)N} \left( 1+ \frac{N^{q}\mu^{q}}{\bar{J} q!}\right).\label{zaveragesaddle2}
\end{align}
We find that the summation of the contributions from  $q$ saddles can recover the exact result \eqref{zaveragesaddle2} by expanding \eqref{zaveragesaddle} up to the order $q$ to find the first nontrivial perturbation.

\subsection{$\langle z\rangle^{2}$}

We start with the partition function \eqref{z2averagemu}. With the similar redefinition \eqref{gsredefinition} and variation over $\sigma_{ab},g_{ab}$  we have the saddle point equations 
\begin{align}
&	\frac{(\mu+\mi \me^{-\frac{\mi\pi}{q}}\sigma_{RR})\me^{-\frac{\mi\pi}{q}}}{(\mu+\mi \me^{-\frac{\mi\pi}{q}}\sigma_{LL})(\mu+\mi \me^{-\frac{\mi\pi}{q}}\sigma_{RR})+ \me^{-\frac{2\mi\pi}{q}}\sigma_{LR}\sigma_{RL}}=g_{LL}, \\
&	\frac{(\mu+\mi \me^{-\frac{\mi\pi}{q}}\sigma_{LL})\me^{-\frac{\mi\pi}{q}}}{(\mu+\mi \me^{-\frac{\mi\pi}{q}}\sigma_{LL})(\mu+\mi \me^{-\frac{\mi\pi}{q}}\sigma_{RR})+ \me^{-\frac{2\mi\pi}{q}}\sigma_{LR}\sigma_{RL}}=g_{RR}, \\
&	\frac{-\mi \me^{-\frac{2\mi\pi}{q}}\sigma_{RL}}{(\mu+\mi \me^{-\frac{\mi\pi}{q}}\sigma_{LL})(\mu+\mi \me^{-\frac{\mi\pi}{q}}\sigma_{RR})+ \me^{-\frac{2\mi\pi}{q}}\sigma_{LR}\sigma_{RL}}=g_{LR}, \\
&	\frac{-\mi \me^{-\frac{2\mi\pi}{q}}\sigma_{LR}}{(\mu+\mi \me^{-\frac{\mi\pi}{q}}\sigma_{LL})(\mu+\mi \me^{-\frac{\mi\pi}{q}}\sigma_{RR})+ \me^{-\frac{2\mi\pi}{q}}\sigma_{LR}\sigma_{RL}}=g_{RL}, \\
&	 \mi\sigma_{LL}+\bar{J}g_{LL}^{q-1}=0,\quad \mi\sigma_{LR}+\bar{J}g_{LR}^{q/2-1}g_{RL}^{q/2}=0, \\
&  \mi\sigma_{RR}+\bar{J}g_{RR}^{q-1}=0, \quad  \mi\sigma_{RL}+\bar{J}g_{LR}^{q/2}g_{RL}^{q/2-1}=0. \label{z2musdeqs}
\end{align}
The above equations have many solutions, but we only pay attention to the two special kinds of saddles which are related to the two contributions in \eqref{z2dominate}.

The first kind of saddles are the diagonal cases where $\sigma_{LR}=\sigma_{RL}=g_{LR}=g_{RL}=0$, then the above equations are two copies of \eqref{zaverageeq}. These saddles are identified as disks, whose contributions can be written as $\langle z\rangle^{2}$.

The second kind of saddles are the non-diagonal cases which are more complicated than the first kind. When $\mu=0$ we can let $\sigma_{LL}=\sigma_{RR}=g_{LL}=g_{RR}=0$ and the remaining equations lead to 
\begin{align}
	\bar{J}g_{RL}^{q/2}g_{LR}^{q/2}+1=0, \label{lrsaddleeqs}
\end{align}
which does not solve all the functions. Recall that from the definition we have $G_{LR}^{\dagger}=G_{RL}$ so that $g_{RL}=\me^{-2\mi\pi/q}g_{LR}^{\dagger}$, to solve the equation we set $g_{LR}=g\me^{\mi\theta},g_{RL}=g\me^{-\mi\theta-2\mi\pi/q}$ where $g,\theta$ are real.  Then the above equation becomes
\begin{align}
		\bar{J}g^{q}-1=0,
\end{align}
and the solution is 
\begin{align}
	g=\bar{J}^{-\frac{1}{q}}\me^{\frac{\mi2m\pi }{q}},\quad m=0,1,\dots q-1. \label{mu0z2solu}
\end{align}
Note that there is a factor $\me^{\mi\theta},\theta\in [0,2\pi)$ so that the phase factor in \eqref{mu0z2solu} is not important. From equation \eqref{lrsaddleeqs} and the conjugation relation, there should be $\frac{q}{2}$ different saddles which are identified as wormholes. 

Then the exponent becomes $\bar{J}^{2N/q}\me^{2N\(\frac{1}{q}-1\)}$, to get the whole result there are several factors. The integral of the one-loop contribution around the saddle is actually three-dimensional due to the conjugation relation between $G_{LR}$ and $G_{RL}$, which gives $\sqrt{4\pi N/q}$ with the measure and the integration over $\theta$. Combining the contribution of $\frac{q}{2}$ saddles, it gives the result $ \langle \Phi^{2}(\mi\me^{\mi\pi/q}\mu) \rangle$ in \eqref{z2dominate}
\begin{align}
	 \frac{(N!)^{2}}{N^{2N}}\left(\frac{N\bar{J}}{q} \right)^{\frac{2N}{q}}  \frac{2^{\frac{2N}{q}}}{\frac{2N}{q}!}\approx \sqrt{\pi Nq}\bar{J}^{2N/q}\me^{2N\left( \frac{1}{q}-1\right)} .   \label{whcontri1}
\end{align}  

Then we consider the small $\mu$ perturbation in equations \eqref{z2musdeqs} for the non-diagonal case.  Now we let $\sigma_{LL}=\sigma_{RR}=0,g_{LL}=g_{RR}=\frac{\mu\me^{-\mi\pi/q}}{\mu^{2}+ \me^{-2\mi\pi /q}\sigma_{LR}\sigma_{RL}}$ then the equations lead to 
\begin{align}
	\mu^{2}\me^{2\mi\pi/q} g_{LR}g_{RL}-\bar{J}^{2}g_{LR}^{q}g_{RL}^{q}=\bar{J} g_{LR}^{q/2}g_{RL}^{q/2}.
\end{align}
The inclusion of $\mu$ does not help for solving all the functions, we still need the conjugation relation between $G_{LR}$ and $G_{RL}$ which means we have  $g_{LR}=g\me^{\mi\theta},g_{RL}=g\me^{-\mi\theta-2\mi\pi/q}$ just like before. Then the equation becomes 
\begin{align}
	\mu^{2} g^{2}-\bar{J}^{2}g^{2q}=-\bar{J} g^{q},
\end{align}
 taking the ansatz $g=g_{0}+g_{1}$ and we have the solution 
\begin{align}
	\bar{J}g_{0}^{q}=1,\quad g_{1}=\frac{\mu^{2}}{q}g_{0}^{3}.
\end{align}
Inserting the solution into the partition function we have 
\begin{align}
	\sqrt{\pi Nq}\bar{J}^{2N/q}\me^{2N\(\frac{1}{q}-1\)-N\mu^{2}J^{-2/q}\me^{4\mi m\pi/q}},\label{whcontri12}
\end{align}
where each $m$ has different contributions. But by expanding the contribution up to the order $\frac{q}{2}$ and summing over $\frac{q}{2}$ saddles, it recovers the exact result which is $n_{3}=\frac{2N}{q},\frac{2N}{q}-1,n_{1}=n_{2}=0$ in \eqref{muz2average}
\begin{align}
	\sqrt{\pi N q}\bar{J}^{2N/q} \me^{2N\(\frac{1}{q}-1\)}\(1+ \frac{(-1)^{q/2}\mu^{q}N^{q/2}}{\bar{J}(q/2)!} \).  \label{whcontri2}
\end{align}
Comparing \eqref{zaveragesaddle},\eqref{zaveragesaddle2} and \eqref{whcontri12},\eqref{whcontri2}  we can find that the disk gets more enhancements from small $\mu$ than the wormhole, which is similar to \cite{Saad:2021rcu}.

\section{Requirements of half-wormholes}

In \cite{Saad:2021rcu},  they have the half-wormhole approximation $z^{2}\approx \langle z^{2}\rangle+\Phi(0)$. Obviously $\Phi(0)$ represents the non-self-averaged part of $z^{2}$ so the first requirement of it is that the average is zero $\langle \Phi(0)\rangle=0$. Then we consider the norm of it by squaring the approximation and taking the average 
\begin{align}
	\langle z^{4}\rangle\approx \langle z^{2}\rangle^{2}+\langle \Phi(0)^{2}\rangle,
\end{align}
which means the norm $\langle \Phi(0)^{2}\rangle$ represents  the variance $\langle z^{4}\rangle-\langle z^{2}\rangle^{2}$. 
Therefore the half-wormholes should have zero mean and whose norm gives the variance of the quantity to be approximated. So we can first find out the dominant contributions of the second moment of the quantity, and divide them into the mean and the variance. We can construct the half-wormhole according to the form of the variance.

In our case the dominant terms of the second moment depend on $\bar{J},\mu$. When $\mu$ is large compared to $\bar{J}$ only the mean dominates, after removing the mean it seems that the left dominant part is very complicated which is similar to the case with large $u$ in \cite{Peng:2022pfa}. When $\mu$ is very small compared to $\bar{J}$ the dominant parts are shown in \eqref{muz2approxzp},\eqref{muz4approxzp}, the half-wormholes can be constructed following \cite{Saad:2021rcu} and we will have some computations in this case later. And it seems there is no other case in the 0-dimensional complex SYK model just like in Fig. \ref{z2distribut}. It is difficult to  analytically get the  transition point $\mu^{\ast}$  since we have to solve the equation $\langle z\rangle^{2}\approx\langle \Phi^{2}(\mi\me^{\mi\pi/q}\mu) \rangle$, numerically we may get the $\mu^{\ast}$  for particular $N,q,\bar{J}$.

\section{Fixed couplings with small $\mu$ }

Now we consider  the half-wormhole contribution in the fixed-coupling theory, but only with small $\mu$ as argued in the above. The  half-wormholes in $z$  and $z^{2}$ are called unlinked half-wormholes and linked half-wormholes respectively, which is explained in \cite{Saad:2021rcu,Peng:2022pfa}.

\begin{figure}[htbp]
	\centering
	\subcaptionbox{}{\includegraphics[scale=0.5]{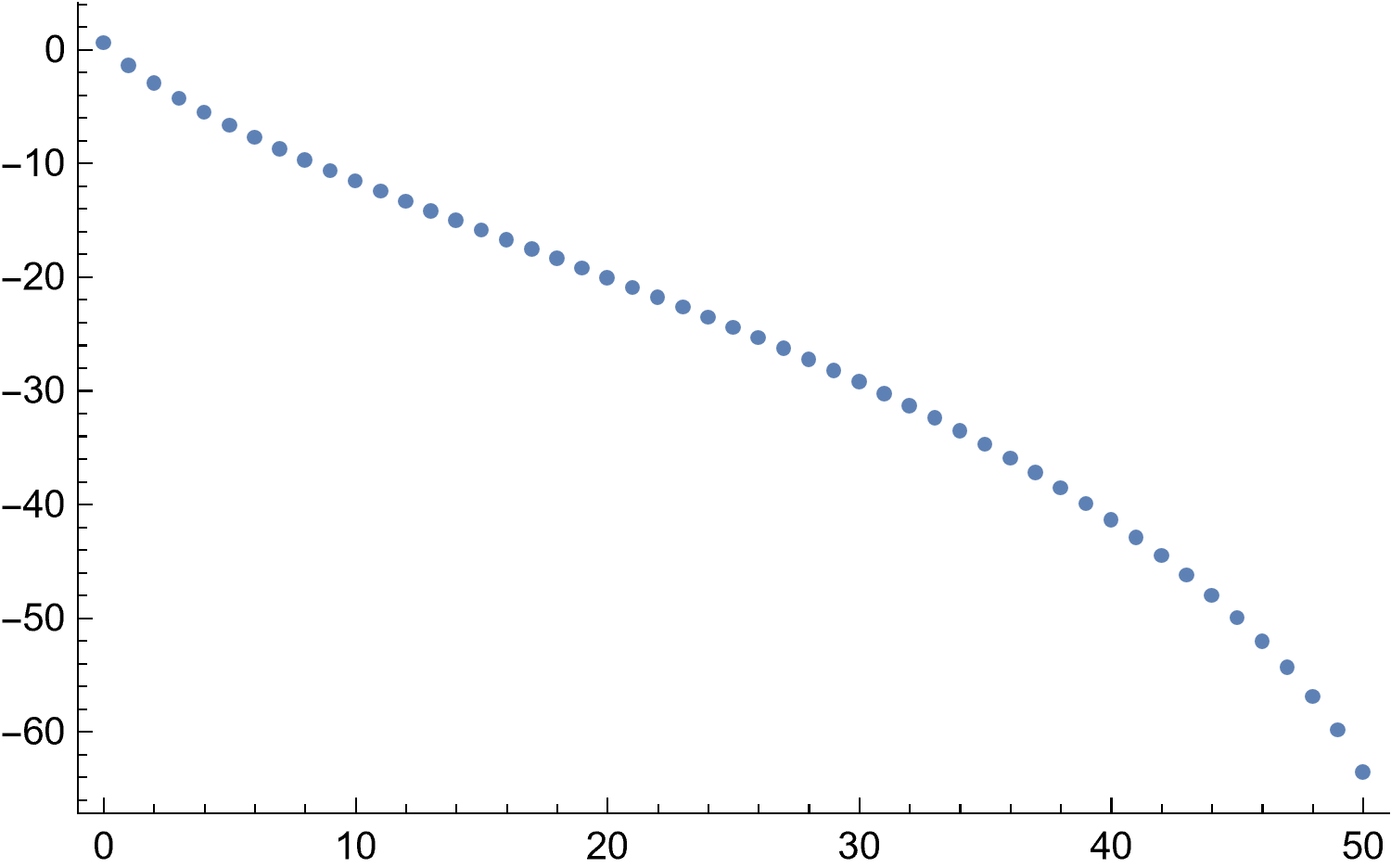}}
	\hfill
	\subcaptionbox{}{\includegraphics[scale=0.5]{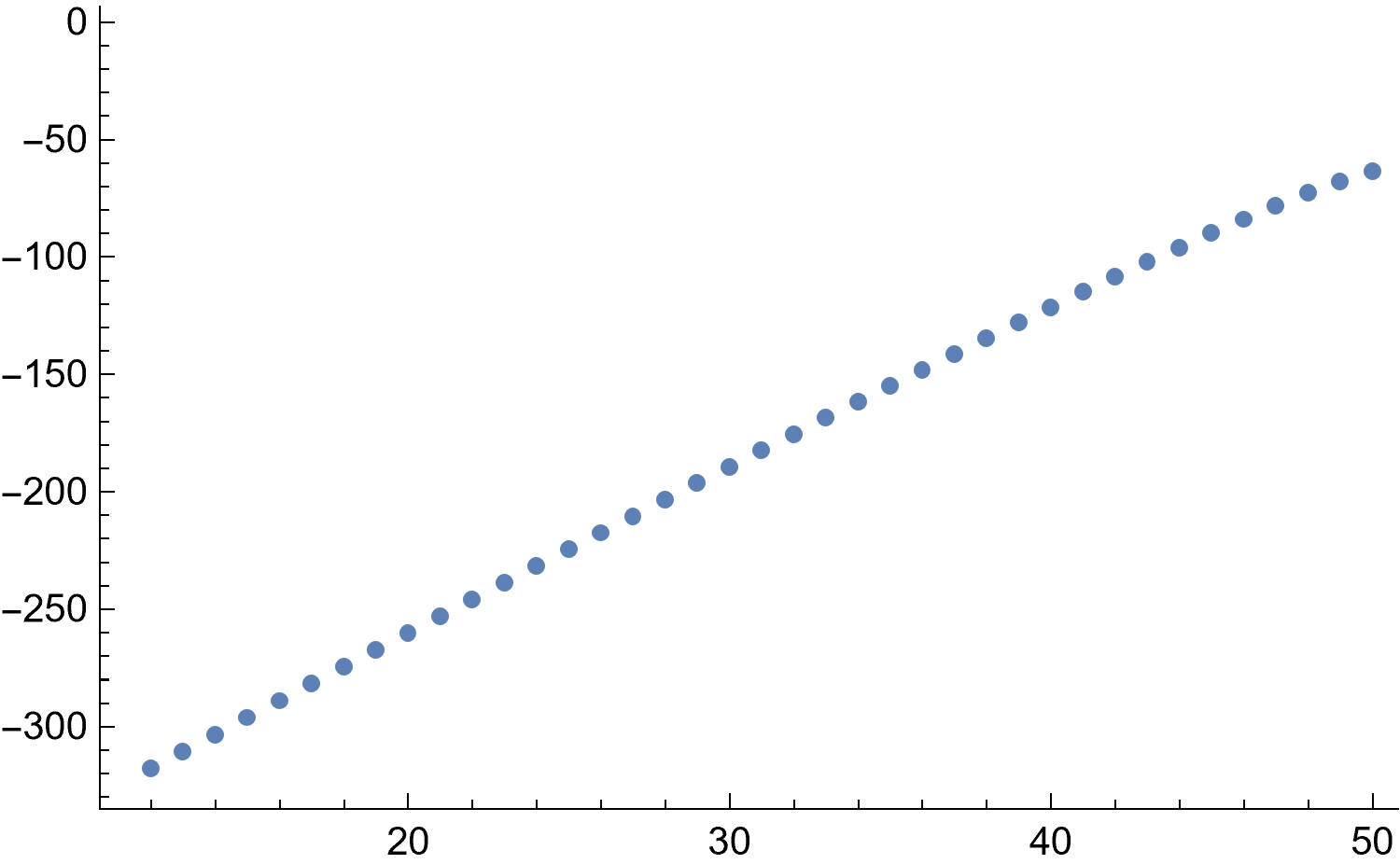}}
	\caption{The absolute values of the terms in the common logarithm $\log_{10}$ in the summation \eqref{muzaverage} with $n$ as the horizontal axis  and  $N=100,q=4,\bar{J}=1,\mu=0.01$. (a) $\sigma=1$. (b) $\sigma=\mi\me^{\mi\pi/q}\mu+0.01$.}
	\label{phidistribut}
\end{figure}

\subsection{$z$}

Following the  procedure in \cite{Saad:2021rcu}, $z$ can be written as a product 
\begin{align}
	z=\int\md\sigma \Psi(\sigma)\Phi(\sigma),
\end{align}
where 
\begin{align}
	\Psi(\sigma)&=\int_{\mathbb{R}}\frac{\md g}{2\pi/N}\exp\left[ N\left( -\mi\sigma g-\frac{\bar{J}}{q}g^{q} \right) \right], \label{psisigma}\\
	\Phi(\sigma)&=\int  \md\psi \md\psi^{\dagger} \exp\left[ \left( \mu +\mi\me^{-\frac{\mi\pi}{q}}\sigma \right) \sum_{i=1}^{N}\psi_{i}^{\dagger}\psi_{i}   \right. \nonumber\\
	&\left.\quad -J_{j_{1}\dots j_{q/2},k_{1}\dots k_{q/2}}\psi^{\dagger}_{j_{1}}\dots \psi^{\dagger}_{j_{q/2}} \psi_{k_{1}}\dots \psi_{k_{q/2}} -\frac{N\bar{J}}{q} \left(\frac{1}{N}\sum_{i=1}^{N}\psi_{i}^{\dagger}\psi_{i}\right)^{q} \right].
\end{align} 
Taking the ensemble average over the coupling in $\Phi(\sigma)$
\begin{align}
	\langle \Phi(\sigma)\rangle=\left( \mu+ \mi\me^{-\mi\pi/q}\sigma\right)^{N},
\end{align}
we will recover the computation of $\langle z\rangle$.  We expect the fluctuation $z-\langle z\rangle$ is captured by the quantity $\Phi\left(\mi\me^{\mi\pi/q}\mu\right)$ since it vanishes after ensemble average. To verify it we consider the quantity $	\langle \Phi^{2}(\sigma) \rangle$ and introduce $G,\Sigma$ fields to rewrite it as 
\begin{align}
&	\langle \Phi^{2}(\sigma) \rangle=\int_{\mathbb{R}\times \mi\mathbb{R} }\frac{\md G_{ab}\md \Sigma_{ab}}{(2\pi\mi/N)^{2}} \exp\left[ N\log\det \left(\Sigma_{ab} \right) \right.\nn\\
	&\left.\quad -N  \Sigma_{LR} G_{LR}-N \Sigma_{RL} G_{RL}+ \frac{2N\bar{J}}{q}  G_{LR}^{q/2}G_{RL}^{q/2}  \right], \label{phi2aveaction}
\end{align}
where $a,b=L,R$ and $\Sigma_{LL}=\Sigma_{RR}=\mu+\mi\me^{-\mi\pi/q}\sigma$.  
By some direct computation we have 
\begin{align}
	\langle \Phi^{2}(\sigma) \rangle&=N!\sum_{m+nq/2=N,m,n\geq 0} \frac{1}{N^{qn}}\left(\frac{2N\bar{J}}{q} \right)^{n} \frac{\left( \mu +\mi\me^{-\frac{\mi\pi}{q}}\sigma \right)^{2m}(-1)^{qn/2}\frac{qn}{2}!}{m!n!}, \label{muphi2average}
\end{align}
where $m,n$ are integers. Like in Fig. \ref{phidistribut}  when $\left( \mu +\mi\me^{-\mi\pi/q}\sigma \right)$ is large the dominant contribution of the above is $n=0$, which gives $\langle \Phi^{2}(\sigma) \rangle=\langle \Phi(\sigma)\rangle^{2}=\left( \mu +\mi\me^{-\mi\pi/q}\sigma \right)^{2N}$, so it is the self-averaged region. When $\sigma$ is around $\mu$ like $\sigma=\mi\me^{\mi\pi/q}\mu$, obviously  $\langle \Phi^{2}(\mi\me^{\mi\pi/q}\mu) \rangle\neq\langle \Phi(\mi\me^{\mi\pi/q}\mu)\rangle^{2}=0$ and the dominant part is $n=\frac{2N}{q}$, which is the non-self-averaged region.

So we propose the approximation from the dominant parts of  $\langle z^{2}\rangle$ which is the second moment of $z$ 
\begin{align}
	z\approx\langle z\rangle + \Phi(\mi\me^{\mi\pi/q}\mu). \label{muapproxi}
\end{align}
 We can  take a saddle point analysis for \eqref{phi2aveaction}, but it is almost the same to the non-diagonal case of $\langle z^{2}\rangle$ so we will not repeat it here.  By the saddle point analysis around \eqref{whcontri1} the term $\langle \Phi^{2}(\mi\me^{\mi\pi/q}\mu) \rangle$ is the same to the wormhole contribution therefore it can be  identified as a wormhole. Similarly from \eqref{whcontri2} $\langle \Phi^{2}(0) \rangle$ is identified as a wormhole plus the small $\mu$ perturbation. Therefore from the argument in \cite{Saad:2021rcu,Peng:2022pfa} $\Phi(\mi\me^{\mi\pi/q}\mu)$ in \eqref{muapproxi} is identified as a  half-wormhole and $\langle z\rangle$ is identified as a disk.

 We define an error to diagnose the approximation 
\begin{align}
	\text{Error}=z-\langle z\rangle -\Phi(\mi\me^{\mi\pi/q}\mu)
\end{align}
whose averages are $\langle \text{Error}\rangle=0$ and
\begin{align}
	\langle \text{Error}^{2}\rangle&=\langle z^{2} \rangle + \left\langle \Phi(\mi\me^{\mi\pi/q}\mu)^{2} \right\rangle-\langle z \rangle^{2}-2\left\langle z\Phi(\mi\me^{\mi\pi/q}\mu)\right \rangle. \label{muzerror}
\end{align}
The only unknown quantity is $\left\langle z\Phi(\mi\me^{\mi\pi/q}\mu)\right \rangle$ which can be similarly computed
\begin{align}
&	\left\langle z\Phi(\mi\me^{\mi\pi/q}\mu) \right\rangle=\int_{\mathbb{R}\times \mi\mathbb{R}} \frac{\md G_{ab}\md \Sigma_{ab}}{(2\pi\mi/N)^{2}} \exp\left[ N\log\det \left(\Sigma_{ab} \right) - N \Sigma_{LL}G_{LL}-N  \Sigma_{LR} G_{LR} \right.\nonumber\\
	&\left.\quad -N \Sigma_{RL} G_{RL}  +  \frac{N}{q}  G_{LL}^{q}+\frac{2N\bar{J}}{q}  G_{LR}^{q/2}G_{RL}^{q/2}  \right],
\end{align}
where $\Sigma_{RR}=0$, and the exact result is 
\begin{align}
\left	\langle z\Phi(\mi\me^{\mi\pi/q}\mu) \right\rangle= \frac{(N!)^{2}(2N\bar{J}/q)^{2N/q}}{N^{2N}(2N/q)!}. \label{muzphi0average}
\end{align}
Then we find the dominant part of the error vanishes $\frac{	\langle \text{Error}^{2}\rangle}{\langle z^{2} \rangle} \ll 1 $, which means the approximation \eqref{muapproxi} is good.

\subsection{$z^{2}$}

Similarly $z^{2}$ can be written as a product 
\begin{align}
	z^{2}=\int\md\sigma_{ab} \Theta(\sigma_{ab})\Lambda(\sigma_{ab}),
\end{align}
where 
\begin{align}
	\Theta(\sigma_{ab})&=\int_{\mathbb{R}}\frac{\md g_{ab}}{(2\pi/N)^{2}}\exp\left[ N\left( -\mi\sigma_{ab} g_{ab}-\frac{\bar{J}}{q}g_{ab}^{q/2}g_{ba}^{q/2} \right) \right],\nn\\
	\Lambda(\sigma_{ab})&=\int  \md\psi^{a} \md\psi^{a\dagger} \exp\left[ \left(\mu\delta_{ab}+ \mi\me^{-\mi\pi/q}\sigma_{ab} \right)\sum_{i=1}^{N}\psi_{i}^{a\dagger}\psi_{i}^{b}  -J_{j_{1}\dots j_{q/2},k_{1}\dots k_{q/2}}\psi^{a\dagger}_{j_{1}}\dots \psi^{a\dagger}_{j_{q/2}} \psi_{k_{1}}^{a}\dots \psi_{k_{q/2}}^{a} \right.\nonumber\\
	&\left.\quad-\frac{N\bar{J}}{q} \left(\frac{1}{N}\sum_{i=1}^{N}\psi_{i}^{a\dagger}\psi_{i}^{b}\right)^{q/2}\left(\frac{1}{N}\sum_{i=1}^{N}\psi_{i}^{b\dagger}\psi_{i}^{a}\right)^{q/2} \right],
\end{align} 
and $a,b=L,R$. If we take the ensemble average over the coupling 
\begin{align}
	\langle \Lambda(\sigma_{ab}) \rangle=\det\(\mu\delta_{ab}+\mi\me^{-\mi\pi/q}\sigma_{ab}\)^{N},
\end{align}
we will recover the computation of $\langle z^{2}\rangle$. We expect that the difference $z^{2}-\langle z^{2}\rangle$ is captured by the $\Lambda$ function so we  consider $	\langle \Lambda(\sigma_{ab})^{2} \rangle$
\begin{align}
	\langle \Lambda(\sigma_{ab})^{2} \rangle&=\int_{\mathbb{R}\times \mi\mathbb{R}} \frac{\md G_{\bar{a}\bar{b}}\md \Sigma_{\bar{a}\bar{b}}}{(2\pi\mi/N)^{8}} \exp\left[ N\log\det \left(\mu\delta_{\bar{a}\bar{b}}+\Sigma_{\bar{a}\bar{b}} \right) \right.\nn\\
	&\left.\quad-N  \Sigma_{\bar{a}\bar{b}} G_{\bar{a}\bar{b}}+ \frac{N\bar{J}}{q}  G_{\bar{a}\bar{b}}^{q/2}G_{\bar{b}\bar{a}}^{q/2}  \right], \label{mulambdasigmasquare}
\end{align}
where in the latter two terms $\bar{a}\bar{b}=L\bar{L},L\bar{R},R\bar{L},R\bar{R},\bar{L}L,\bar{L}R,\bar{R}L,\bar{R}R$. In the determinant except the previous values,  $\bar{a}\bar{b}$ can take all the other values and in these cases $\Sigma_{\bar{a}\bar{b}}=\mu\delta_{\bar{a}\bar{b}}+\mi\me^{-\mi\pi/q}\sigma_{ab},a,b=L,R $ in these values. The matrix of the derivative is 
\begin{align}
	\left(
	\begin{array}{cccc}
		\mu+\sigma_{LL} &\sigma_{LR}&  \partial_{G_{L\bar{L}}} &  \partial_{G_{L\bar{R}}}\\
		\sigma_{RL} &\mu+\sigma_{RR} &  \partial_{G_{R\bar{L}}}& \partial_{G_{R\bar{R}}}\\
		\partial_{G_{\bar{L}L}} & \partial_{G_{\bar{L}R}} & \mu+ \sigma_{LL}& \sigma_{LR}\\
			\partial_{G_{\bar{R}L}} & \partial_{G_{\bar{R}R}} & \sigma_{RL}& \mu+\sigma_{RR}\\
	\end{array}
	\right) , \label{mulambda2matrix}
\end{align}
where we neglect the factors $N,\mi\me^{-\mi\pi/q}$ for simplicity.  We can obtain the dominant terms following the computation of $\langle z^{4}\rangle$ \eqref{z4aveaction}.
The self-averaged region  is located at large $\sigma$ which makes  $\langle \Lambda(\sigma_{ab})^{2} \rangle =	\langle \Lambda(\sigma_{ab}) \rangle^{2}=\det(\mu\delta_{ab}+\mi\me^{-\mi\pi/q}\sigma_{ab})^{2N}$.  While when $\sigma_{ab}$ locate at the below points 
\begin{align}
	\bar{\sigma}_{LL}=\mi\me^{\mi\pi/q}\mu,\quad 	\bar{\sigma}_{LR}=0,\quad 	\bar{\sigma}_{RL}=0,\quad 		\bar{\sigma}_{RR}=\mi\me^{\mi\pi/q}\mu,
\end{align}
$\langle \Lambda(	\bar{\sigma}_{ab})^{2} \rangle \neq \langle \Lambda(\bar{\sigma}_{ab}) \rangle^{2}=0$, which is the non-self-averaged region.

Therefore we propose an approximation
\begin{align}
	\left( z-\langle z\rangle \right)^{2}&\approx  \left\langle 	\left( z-\langle z\rangle \right)^{2} \right\rangle+ \Lambda(\bar{\sigma}_{ab}), \label{munewapprox}\\
	z^{2}& \approx \langle z^{2}\rangle + 2z\langle z\rangle -2\langle z\rangle^{2}+\Lambda(\bar{\sigma}_{ab}), \label{muz2approximation2}
\end{align}
where we expand the first row to get the second row.
And the error will be
\begin{align}
	\text{Error}&= 	z^{2} - \langle z^{2}\rangle - 2z\langle z\rangle +2\langle z\rangle^{2}-\Lambda(\bar{\sigma}_{ab}),\\
	\langle \text{Error}\rangle&=0,\\
	\langle \text{Error}^{2}\rangle&=\langle z^{4} \rangle + \langle \Lambda(\bar{\sigma}_{ab})^{2} \rangle-\langle z^{2} \rangle^{2}-2\langle z^{2}\Lambda(\bar{\sigma}_{ab}) \rangle \nn\\
	&\quad+8\langle z^{2}\rangle \langle z\rangle^{2}-4\langle z\rangle^{4}-4\langle z^{3}\rangle\langle z\rangle.
\end{align}
Similarly to solve the above quantity we have to compute  $\langle z^{2} \Lambda(\bar{\sigma}_{ab}) \rangle$ and $\langle z^{3}\rangle$. They  can be evaluated similarly by the determinant of the matrix of the derivative, which gives 
\begin{align}
	\langle \Lambda(\bar{\sigma}_{ab})^{2} \rangle&\approx\left\langle z^{2} \Lambda(\bar{\sigma}_{ab})\right \rangle \approx 2\left\langle \Phi(\mi\me^{\mi\pi/q}\mu)^{2}\right\rangle^{2}, \label{mulam2approxzp} \\
	\langle z^{3}\rangle&\approx  \langle z\rangle^{3}+3\langle z\rangle \left\langle \Phi(\mi\me^{\mi\pi/q}\mu)^{2}\right\rangle.  \label{muz3approxzp}
\end{align}
Using the relations \eqref{muz2approxzp},\eqref{muz4approxzp},\eqref{mulam2approxzp},\eqref{muz3approxzp} we find the dominant part vanishes $	\frac{	\langle \text{Error}^{2}\rangle}{\langle z^{4} \rangle} \ll 1 $, which means the approximation \eqref{muz2approximation2} is good. 

For $\langle z^{4}\rangle$ and $\langle \Lambda(\sigma_{ab})^{2} \rangle$ it is similar to take the saddle point analysis and get similar conclusions,  here we omit it due to its computational complexity. The approximation \eqref{munewapprox} means the sum of a wormhole and a half-wormhole can recover the factorization.

\section{Conclusion} \label{conclusion}

In this paper  we identify the half-wormhole contribution in a 0-dimensional complex SYK model by exactly evaluating the Grassmann integral and the saddle point analysis. The chemical potential $\mu$ is related to the strength of the bulk gauge field, which can effectively modify the computation and the relative magnitude of the half-wormhole.  When $\mu=0$ the computation is  similar but not the same to two decoupled Majorana SYK models and the saddle point analysis is also available. When $\mu$ is nonzero the saddle point analysis is difficult, we can use the exact computation or perturbative saddle point analysis.  When $\mu$ is very small, perturbatively the disk  receives more enhancement from nonzero $\mu$ to the wormhole and  the half-wormhole. When $\mu$ is large compared to $\bar{J}$ the disk dominates, approximately  there is no wormhole and half-wormhole contribution. 

For future directions it is interesting to  find other ways to construct the half-wormhole, which is applicable to more general cases such as finite $\mu$ in our theory. It is also interesting to consider the bulk duals of the half-wormholes or identify the half-wormhole contributions in other models.

\section*{Acknowledgments}
We thank Cheng Peng for the helpful advice.  This research is supported in part by NSFC Grant No.11735001,12275004.


\begin{thebibliography}{10}
	
	\small
	
	\bibitem{Almheiri:2019qdq}
	A.~Almheiri, T.~Hartman, J.~Maldacena, E.~Shaghoulian and A.~Tajdini,
	``Replica Wormholes and the Entropy of Hawking Radiation,''
	JHEP \textbf{05}, 013 (2020)
	doi:10.1007/JHEP05(2020)013
	[arXiv:1911.12333 [hep-th]].
	

    \bibitem{Penington:2019kki}
    G.~Penington, S.~H.~Shenker, D.~Stanford and Z.~Yang,
   ``Replica wormholes and the black hole interior,''
    JHEP \textbf{03}, 205 (2022)
    doi:10.1007/JHEP03(2022)205
    [arXiv:1911.11977 [hep-th]].
	
	\bibitem{Saad:2019lba}
	P.~Saad, S.~H.~Shenker and D.~Stanford,
	``JT gravity as a matrix integral,''
	[arXiv:1903.11115 [hep-th]].
	
	\bibitem{Saad:2018bqo}
	P.~Saad, S.~H.~Shenker and D.~Stanford,
	``A semiclassical ramp in SYK and in gravity,''
	[arXiv:1806.06840 [hep-th]].
	
	\bibitem{Saad:2019pqd}
	P.~Saad,
	``Late Time Correlation Functions, Baby Universes, and ETH in JT Gravity,''
	[arXiv:1910.10311 [hep-th]].
	
	
	
	\bibitem{Maldacena:1997re}
	J.~M.~Maldacena,
	``The Large N limit of superconformal field theories and supergravity,''
	Adv. Theor. Math. Phys. \textbf{2}, 231-252 (1998)
	doi:10.1023/A:1026654312961
	[arXiv:hep-th/9711200 [hep-th]].
	
	\bibitem{Witten:1998qj}
	E.~Witten,
	``Anti-de Sitter space and holography,''
	Adv. Theor. Math. Phys. \textbf{2}, 253-291 (1998)
	doi:10.4310/ATMP.1998.v2.n2.a2
	[arXiv:hep-th/9802150 [hep-th]].
	
	\bibitem{Gubser:1998bc}
	S.~S.~Gubser, I.~R.~Klebanov and A.~M.~Polyakov,
	``Gauge theory correlators from noncritical string theory,''
	Phys. Lett. B \textbf{428}, 105-114 (1998)
	doi:10.1016/S0370-2693(98)00377-3
	[arXiv:hep-th/9802109 [hep-th]].
	
	\bibitem{Maldacena:2004rf}
	J.~M.~Maldacena and L.~Maoz,
	``Wormholes in AdS,''
	JHEP \textbf{02}, 053 (2004)
	doi:10.1088/1126-6708/2004/02/053
	[arXiv:hep-th/0401024 [hep-th]].
	
	\bibitem{Coleman:1988cy}
	S.~R.~Coleman,
	``Black Holes as Red Herrings: Topological Fluctuations and the Loss of Quantum Coherence,''
	Nucl. Phys. B \textbf{307}, 867-882 (1988)
	doi:10.1016/0550-3213(88)90110-1
	\bibitem{Giddings:1988wv}
	S.~B.~Giddings and A.~Strominger,
	``Baby Universes, Third Quantization and the Cosmological Constant,''
	Nucl. Phys. B \textbf{321}, 481-508 (1989)
	doi:10.1016/0550-3213(89)90353-2
	\bibitem{Giddings:1988cx}
	S.~B.~Giddings and A.~Strominger,
	``Loss of Incoherence and Determination of Coupling Constants in Quantum Gravity,''
	Nucl. Phys. B \textbf{307}, 854-866 (1988)
	doi:10.1016/0550-3213(88)90109-5
	
	\bibitem{Jackiw:1984je}
	R.~Jackiw,
	``Lower Dimensional Gravity,''
	Nucl. Phys. B \textbf{252}, 343-356 (1985)
	doi:10.1016/0550-3213(85)90448-1
	\bibitem{Teitelboim:1983ux}
	C.~Teitelboim,
	``Gravitation and Hamiltonian Structure in Two Space-Time Dimensions,''
	Phys. Lett. B \textbf{126}, 41-45 (1983)
	doi:10.1016/0370-2693(83)90012-6
	
	\bibitem{Sachdev:1992fk}
	S.~Sachdev and J.~Ye,
	``Gapless spin fluid ground state in a random, quantum Heisenberg magnet,''
	Phys. Rev. Lett. \textbf{70}, 3339 (1993)
	doi:10.1103/PhysRevLett.70.3339
	[arXiv:cond-mat/9212030 [cond-mat]].
	
	\bibitem{KitaevTalk2}
	A. Kitaev, "A simple model of quantum holography." Talks at KITP
	http://online.kitp.ucsb.edu/online/entangled15/kitaev/ and
	http://online.kitp.ucsb.edu/online/entangled15/kitaev2\\
	A. Kitaev, "Hidden correlations in the Hawking radiation and thermal noise." Talk at KITP
	http://online.kitp.ucsb.edu/online/joint98/kitaev/
	
	\bibitem{Maldacena:2016hyu}
	J.~Maldacena and D.~Stanford,
	Phys. Rev. D \textbf{94}, no.10, 106002 (2016)
	doi:10.1103/PhysRevD.94.106002
	[arXiv:1604.07818 [hep-th]].
	
	
	\bibitem{Saad:2021rcu}
	P.~Saad, S.~H.~Shenker, D.~Stanford and S.~Yao,
	``Wormholes without averaging,''
	JHEP \textbf{09}, 133 (2024)
	doi:10.1007/JHEP09(2024)133
	[arXiv:2103.16754 [hep-th]].
	
	\bibitem{Peng:2022pfa}
	C.~Peng, J.~Tian and Y.~Yang,
	``Half-wormholes and ensemble averages,''
	Eur. Phys. J. C \textbf{83}, no.11, 993 (2023)
	doi:10.1140/epjc/s10052-023-12164-9
	[arXiv:2205.01288 [hep-th]].
	
	\bibitem{Forste:2024nsw}
	S.~Forste and S.~Natu,
	``Half-wormholes in a supersymmetric SYK model,''
	JHEP \textbf{02}, 032 (2025)
	doi:10.1007/JHEP02(2025)032
	[arXiv:2411.10155 [hep-th]].
	
	\bibitem{Tian:2022zpc}
	J.~Tian and Y.~Yang,
	``More on half-wormholes and ensemble averages,''
	Commun. Theor. Phys. \textbf{75}, no.9, 095001 (2023)
	doi:10.1088/1572-9494/acde6b
	[arXiv:2211.09398 [hep-th]].
	
	\bibitem{Saad:2021uzi}
	P.~Saad, S.~H.~Shenker and S.~Yao,
	``Comments on wormholes and factorization,''
	JHEP \textbf{10}, 076 (2024)
	doi:10.1007/JHEP10(2024)076
	[arXiv:2107.13130 [hep-th]].
	
	\bibitem{Mukhametzhanov:2021nea}
	B.~Mukhametzhanov,
	``Half-wormholes in SYK with one time point,''
	SciPost Phys. \textbf{12}, no.1, 029 (2022)
	doi:10.21468/SciPostPhys.12.1.029
	[arXiv:2105.08207 [hep-th]].
	
	\bibitem{Garcia-Garcia:2021squ}
	A.~M.~Garc\'\i{}a-Garc\'\i{}a and V.~Godet,
	``Half-wormholes in nearly AdS$_2$ holography,''
	SciPost Phys. \textbf{12}, no.4, 135 (2022)
	doi:10.21468/SciPostPhys.12.4.135
	[arXiv:2107.07720 [hep-th]].
	
	\bibitem{Choudhury:2021nal}
	S.~Choudhury and K.~Shirish,
	``Wormhole calculus without averaging from O(N)q-1 tensor model,''
	Phys. Rev. D \textbf{105}, no.4, 046002 (2022)
	doi:10.1103/PhysRevD.105.046002
	[arXiv:2106.14886 [hep-th]].
	
	
	\bibitem{Mukhametzhanov:2021hdi}
	B.~Mukhametzhanov,
	``Factorization and complex couplings in SYK and in Matrix Models,''
	JHEP \textbf{04}, 122 (2023)
	doi:10.1007/JHEP04(2023)122
	[arXiv:2110.06221 [hep-th]].
	
	\bibitem{Goto:2021mbt}
	K.~Goto, Y.~Kusuki, K.~Tamaoka and T.~Ugajin,
	``Product of random states and spatial (half-)wormholes,''
	JHEP \textbf{10}, 205 (2021)
	doi:10.1007/JHEP10(2021)205
	[arXiv:2108.08308 [hep-th]].
	
	\bibitem{Blommaert:2021fob}
	A.~Blommaert, L.~V.~Iliesiu and J.~Kruthoff,
	``Gravity factorized,''
	JHEP \textbf{09}, 080 (2022)
	doi:10.1007/JHEP09(2022)080
	[arXiv:2111.07863 [hep-th]].
	
	\bibitem{Goto:2021wfs}
	K.~Goto, K.~Suzuki and T.~Ugajin,
	``Factorizing wormholes in a partially disorder-averaged SYK model,''
	JHEP \textbf{09}, 069 (2022)
	doi:10.1007/JHEP09(2022)069
	[arXiv:2111.11705 [hep-th]].
	
	
	\bibitem{Cheng:2022nra}
	P.~Cheng and P.~Mao,
	``Notes on wormhole cancellation and factorization,''
	Eur. Phys. J. C \textbf{84}, no.7, 675 (2024)
	doi:10.1140/epjc/s10052-024-13045-5
	[arXiv:2208.08456 [hep-th]].
	
	\bibitem{Peng:2021vhs}
	C.~Peng, J.~Tian and J.~Yu,
	``Baby universes, ensemble averages and factorizations with matters,''
	[arXiv:2111.14856 [hep-th]].
	
	\bibitem{Gu:2019jub}
	Y.~Gu, A.~Kitaev, S.~Sachdev and G.~Tarnopolsky,
	``Notes on the complex Sachdev-Ye-Kitaev model,''
	JHEP \textbf{02}, 157 (2020)
	doi:10.1007/JHEP02(2020)157
	[arXiv:1910.14099 [hep-th]].
	
	\bibitem{Davison:2016ngz}
	R.~A.~Davison, W.~Fu, A.~Georges, Y.~Gu, K.~Jensen and S.~Sachdev,
	``Thermoelectric transport in disordered metals without quasiparticles: The Sachdev-Ye-Kitaev models and holography,''
	Phys. Rev. B \textbf{95}, no.15, 155131 (2017)
	doi:10.1103/PhysRevB.95.155131
	[arXiv:1612.00849 [cond-mat.str-el]].
	
	
	\bibitem{Gaikwad:2018dfc}
	A.~Gaikwad, L.~K.~Joshi, G.~Mandal and S.~R.~Wadia,
	``Holographic dual to charged SYK from 3D Gravity and Chern-Simons,''
	JHEP \textbf{02}, 033 (2020)
	doi:10.1007/JHEP02(2020)033
	[arXiv:1802.07746 [hep-th]].
	
	\bibitem{Chaturvedi:2020jyy}
	P.~Chaturvedi, I.~Papadimitriou, W.~Song and B.~Yu,
	``AdS$_{3}$ gravity and the complex SYK models,''
	JHEP \textbf{05}, 142 (2021)
	doi:10.1007/JHEP05(2021)142
	[arXiv:2011.10001 [hep-th]].
	
	\bibitem{Afshar:2019axx}
	H.~Afshar, H.~A.~Gonz\'alez, D.~Grumiller and D.~Vassilevich,
	``Flat space holography and the complex Sachdev-Ye-Kitaev model,''
	Phys. Rev. D \textbf{101}, no.8, 086024 (2020)
	doi:10.1103/PhysRevD.101.086024
	[arXiv:1911.05739 [hep-th]].
	
	\bibitem{Godet:2020xpk}
	V.~Godet and C.~Marteau,
	``New boundary conditions for AdS$_{2}$,''
	JHEP \textbf{12}, 020 (2020)
	doi:10.1007/JHEP12(2020)020
	[arXiv:2005.08999 [hep-th]].
	
   \bibitem{Zhang:2025kty}
   Z.~Zhang and C.~Peng,
   ``Gauging the complex SYK model,''
   JHEP \textbf{08}, 217 (2025)
   doi:10.1007/JHEP08(2025)217
   [arXiv:2502.18595 [hep-th]].
	
\end{thebibliography}
\end{document}